\documentclass[letterpaper]{article}

\usepackage{graphicx}
\usepackage{amsmath,amssymb}
\usepackage{color}

\newcommand{\be}{\begin{equation}}
\newcommand{\ee}{\end{equation}\noindent}
\newcommand{\bea}{\begin{eqnarray}}
\newcommand{\eea}{\end{eqnarray}}
\newcommand{\bal}{\begin{align}}

\newcommand{\srnn}{\sqrt{s_{\rm NN}}}

\begin{document}

\title{
Probing QCD Phase Structure using Baryon Multiplicity Distribution
}

\author{Atsushi Nakamura
\thanks{
RIISE, Hiroshima University,
Higashi-Hiroshima 739-8527 Japan.
}
and Keitaro Nagata
\thanks{
KEK, Tsukuba, Ibaraki 305-0801, Japan.
}
}


\date{today}

\maketitle

\begin{abstract}
We propose a new method to construct canonical partition functions of QCD 
from net number distributions
such as the net-baryon, net-charge and net-strangeness, by using only the CP symmetry.
To demonstrate the method, we apply it to the net-proton number distribution 
$P_n$ recently measured at RHIC. 
We show that both $\mu/T$ and the canonical partition functions $Z_n$ 
are determined by using the CP invariance $Z_n =Z_{-n}$. 
Comparing $\mu/T$ obtained from the present analysis for the net-proton distribution 
and that obtained from a thermal statistical model, 
we find remarkable agreement for wide range of beam energies. 
%
%
Constructing a grand canonical partition function $Z(\mu,T)=\sum_n Z_n(T) \xi^n$, 
we study moments and Lee-Yang zeros for RHIC data, and discuss possible regions of a phase transition line in QCD. 
%
This is the first Lee-Yang zero diagram obtained for RHIC data, 
which helps us to see contributions of large net-proton data for exploring
the QCD phase diagram.

%
%
%
 
We also calculate $Z_n$ by the lattice QCD simulations, and
find a clear indication of Roberge-Weiss phase transition in the
QGP phase.
The method does not rely on the Taylor expansions, which prevent us to go to large $\mu/T$.
\end{abstract}

\section{Introduction}

When temperature and density are varied, QCD
is expected to have a rich phase structure \cite{FukushimaHatsuda}.
One of the most important challenges in particle and nuclear
physics is to experimentally discover the
phases that only appear under extreme conditions and 
to theoretically understand their nature.
Such achievements  would not only deepen  our understanding of QCD 
but also extend our knowledge of the early universe and compact stars.

The Relativistic Heavy Ion Collider (RHIC) was built 
to explore the properties of QCD matter
\cite{GyulassyMcLerran}.
Recently net-proton multiplicity measurements at RHIC 
are gaining attention~\cite{Aggarwal:2010wy,STARLuo}
because they provide valuable information about the QCD phase diagram~\cite{Stephanov1998dy}.
In these measurements,
the colliding energy is varied, and
trajectories of the produced hot matter in ($T,\mu_B$) plane 
may pass near the critical region.
Event-by-event fluctuations are expected to encompass the
critical point, where the correlation length rapidly changes
\cite{Friman2011,ARS2010,Stephanov2011}.
In particular, conserved quantities such as the charge or baryon number
may reveal possible correlations that existed inside the system
before hadronization. 
See Ref.\cite{Morita2012} and references therein.

Usually, data obtained at a given colliding energy is assigned with a
set of temperature $T$ and chemical potential $\mu$, which are referred to
as chemical freeze-out point, due to the success of a thermal statistical model 
for hadrons in heavy ion collisions
\cite{BRS2004}.

In this paper, we propose a method by which, we can obtain information 
of the QCD phase diagram not only at the experimental $(T,\mu/T)$ point, but other values of $\mu/T$. 
This may seem to be magical, 
but is possible because we take into account all the multiplicities. 
Basic idea is the use of CP symmetry to extract canonical partition functions 
from measured number distribution according to a fundamental 
equation in statistical mechanics. 
This idea also provides a model independent method to determine 
$\mu/T$ only using CP symmetry. 
As we will show later, $\mu/T$ obtained in the present method is consistent 
with those obtained from the chemical freeze-out data, for wide range of colliding energies.

In addition, even without direct lattice QCD calculations 
in physical chemical potential
regions, we can calculate  the canonical partition functions,
which helps us to  understand the QCD at the finite density. 
The method provides us an approach beyond the Taylor expansion method,
namely it is possible to calculate large $\mu/T$ regions.

In section 2, we describe how to extract the canonical partition
functions, $Z_n$, from  experimental data, and show the obtained results.
In section 3, we construct the grand partition function $Z(\mu,T)$ 
from these $Z_n$, and calculate the moments as a function of $\mu/T$.
In section 4, we show that we can calculate the Lee-Yang zero structure
from the canonical partition functions obtained by the lattice QCD 
and the RHIC experiment.  To our knowledge, this is the first calculation
of the Lee-Yang zero for the high energy nuclear collisions.
Section 5 is devoted to the concluding remarks.
In Appendix, we give detailed numerical data $Z_n$ 
and Moments $\lambda_4/\lambda_2$ for RHIC data.
Part of the results here were reported in several proceedings
\cite{APPC12,QM2014}.

\section{Constructing the canonical partition functions}
\label{Sec:Zn}

The grand partition function $Z$ and the canonical partition
functions $Z_n$ are related as 
\be
Z(\xi,T) = \mbox{Tr}\, e^{-(H-\mu\hat{N})/T}
= \sum_{n=-N_{max}}^{+N_{max}} \langle n|e^{-H/T}|n\rangle e^{\mu n/T}
= \sum_{n=-N_{max}}^{+N_{max}} Z_n(T) \xi^n ,
\label{Eq:GrandCanonical}
\ee
where $\xi=\exp(\mu/T)$ is fugacity, and
\be
Z_n = \langle n|\exp(-H/T)|n \rangle.
\label{Eq:Zn}
\ee
Here we assume that the number operator $\hat{N}$ commutes
with $H$, that is, $\hat{N}$ is a conserved quantity.
$\hat{N}$ can be any conserved number operators, such as baryon, charge and
strangeness.  In the following we consider the baryon number
as a concrete example.

\begin{figure} 
  \centering
 \includegraphics[width=0.7\linewidth]{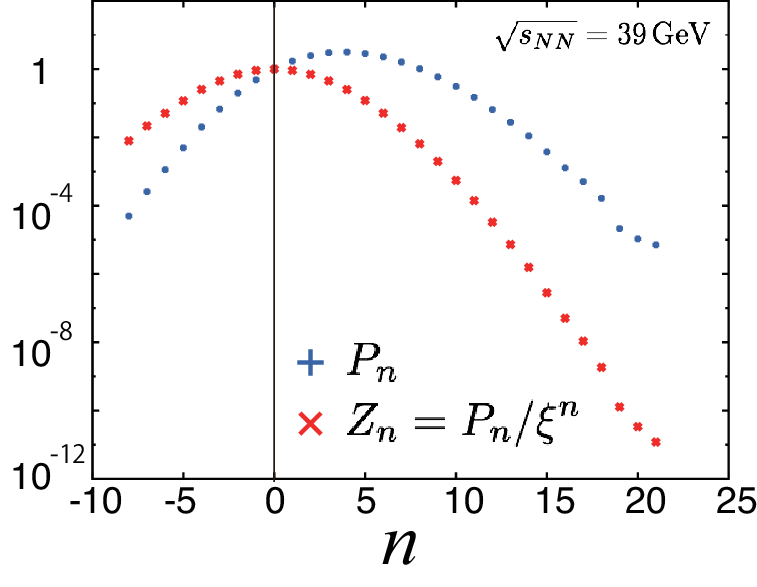}

\caption
{Experimental multiplicity data $P_n$, and 
$Z_n=P_n/\xi^n$ for $\sqrt{s_{NN}}=39$ GeV.
$\xi$ is tuned so that $Z_n=Z_{-n}$, and $\xi$=1.88336.
}
  \label{Fig:PandZ}
\end{figure}

Because of the charge-parity symmetry,
$Z_n$ defined by Eq.(\ref{Eq:Zn}) satisfy
\be
Z_n = Z_{-n} .
\label{Eq:CP}
\ee
The multiplicity distributions $P_n$ observed in experiments
are  related to $Z_n$ as
\be
P_n(\xi) = Z_n \xi^n.
\label{Eq:P-Z}
\ee
Using Eqs.(\ref{Eq:CP}) and (\ref{Eq:P-Z}), we can determine $\xi$
and $Z_n$.
In Fig.\ref{Fig:PandZ} and Table \ref{Tbl:fugacity}, 
we show as an example $P_n(\xi)$ and $Z_n \xi^n$
for $\sqrt{s_{NN}}=39$ GeV\cite{STARLuo}.
The data correspond to the 0-5\% centrality in  Au+Au collisions.
Here $\xi$=1.88336 as given in Table \ref{Tbl:fugacity}.

\begin{figure} 
  \centering
 \includegraphics[width=0.7\linewidth]{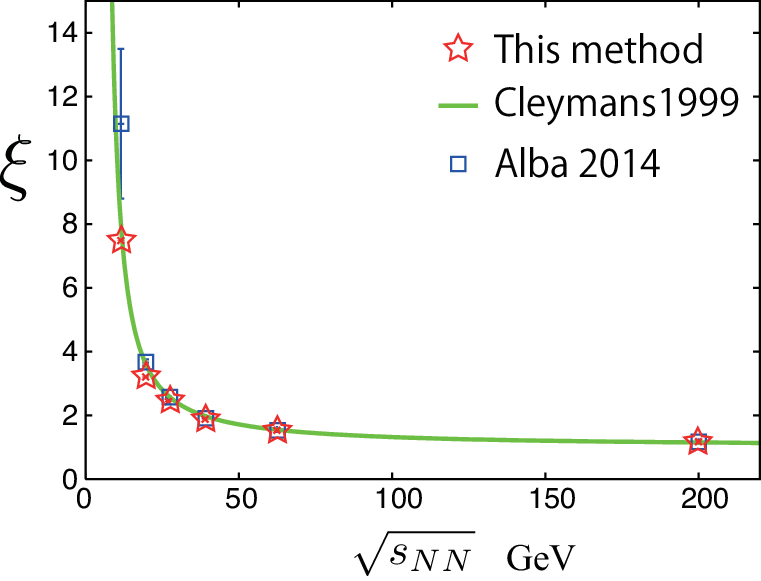}

\caption
{Fugacity $\xi=\exp(\mu/T)$
as a function of the colliding energy $\srnn$.
Plotted values are those obtained from RHIC experiments 
(red star) and the freeze-out results reported
in Refs.\cite{Cleymans2006}(green line) and \cite{Alba2014}(blue square).} 
  \label{Fig:FreezeOut}
\end{figure}

Figure\ref{Fig:FreezeOut} shows the obtained $\xi$ together with
that obtained by freeze-out analysis in Refs.\cite{Cleymans2006} and
\cite{Alba2014}.
Errors in $\xi$ due to the breaking of the relation Eq.(\ref{Eq:CP})
are less than one percent for all colliding energies.
Note that we use here only the multiplicity data and the charge-parity
symmetry relation (\ref{Eq:CP}).

We assume that the net-proton multiplicity data are
approximately proportional to those of the baryon\footnote{
The proportionality factor has no effect on any of the results reported
in this paper.}. 
This approximation is justified if 
(i) after the chemical freeze-out, the net-proton number is effectively constant, 
or
(ii) a created fireball is approximately isoneutral.
See also Sec.3 in Ref.\cite{LuoActPhyPol}.

\begin{table}
\begin{tabular}{cccc}
\hline 
$\sqrt{s_{NN}}$ GeV & $\xi$ obtained here & Freeze-out (Ref.\cite{Cleymans2006})
& Freeze-out (Ref.\cite{Alba2014}) \\
\hline 
  11.5           & 7.48331 $\pm$ 3.85E-6           & 8.040  & 11.1 $\pm$ 2.3 \\
  19.6           & 3.20376 $\pm$ 1.504E-2           & 3.623  & 3.659 $\pm$ 9.7 \\
  27              & 2.43956 $\pm$  5.05E-3          & 2.615 &  2.573 $\pm$ 2.4 \\
  39              & 1.88336 $\pm$  1.18E-3           & 1.981  & 1.936 $\pm$ 1.8\\
  62.4          & 1.53377 $\pm$  2.82E-4           & 1.551  & 1.5573 $\pm$ 6.2 \\
  200           & 1.17499 $\pm$  8.64E-5           & 1.152  &  1.1800 $\pm$ 4.8\\ 
\hline 
\end{tabular}
\caption{The fugacity $\xi$ at each $\sqrt{s_{NN}}$.
The values obtained by the method proposed here are compared with those by the freeze-out analyses.
\label{Tbl:fugacity}
}
\end{table}

\section{Moments as a function of $\mu/T$ and the QCD
transition boundary}

From the grand partition function given by Eq.(\ref{Eq:GrandCanonical}),
the moments are evaluated using 
\be
\lambda_k(\xi) \equiv 
\biggl(\xi\frac{\partial}{\partial\xi}\biggr)^k \log Z(\xi) .
\label{Eq:MomentsDef}
\ee

\begin{figure} 
  \includegraphics[width=1.00\linewidth]{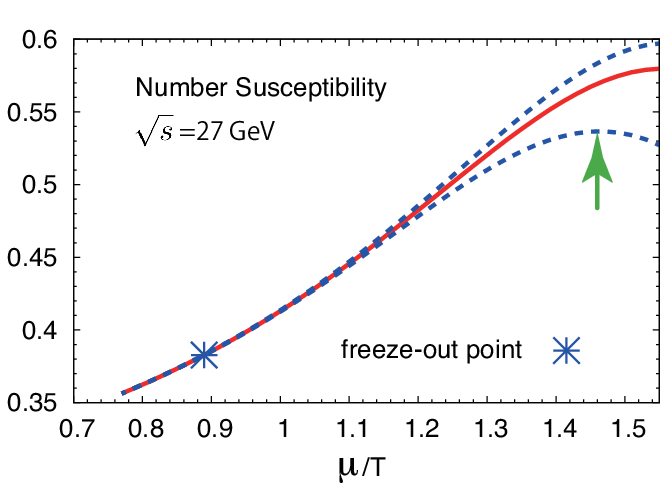}
\caption{
Number susceptibility, $\lambda_2/N_{\rm max}$, at $\srnn=27$ GeV 
as a function of $\mu/T$.
The upper curve is obtained by increasing 
the final three $Z_n ( n = N_{\rm max}-2, \, N_{\rm max}-1, \, N_{\rm max}$)
by 15\%.
The lower curve results from removing the final two
$Z_n$ terms.
\label{Fig:Sus-27}
}
\end{figure}

The quantity $N_{\rm max}$ in Eq.(\ref{Eq:GrandCanonical}) is finite
because of the measurement statistics 
(simulation statistics and finite volume)
in the experiments (lattice QCD). 
The finite nature of $N_{\rm max}$ places an upper bound
on the chemical potential for which the calculation is reliable.
To estimate the effect of the finite $N_{\rm max}$, 
we test two cases:
\begin{enumerate}
\item
the values of the final three $Z_n$
($n=N_{\rm max}-2$,\,$N_{\rm max}-1$,\,$N_{\rm max}$)
are increased by 15\% 
\item
the final two $Z_n$ 
($n=N_{\rm max}-1$,\, $N_{\rm max}$)
are set to zero.
\end{enumerate}
As an example, we plot the number susceptibility $\lambda_2/N_{\rm max}$
in Fig.\ref{Fig:Sus-27} as a function of $\mu/T$
at $\srnn=27$ GeV for these two cases 
together with the one constructed from all the $Z_n$.

\begin{figure} 
  \includegraphics[width=1.00\linewidth]{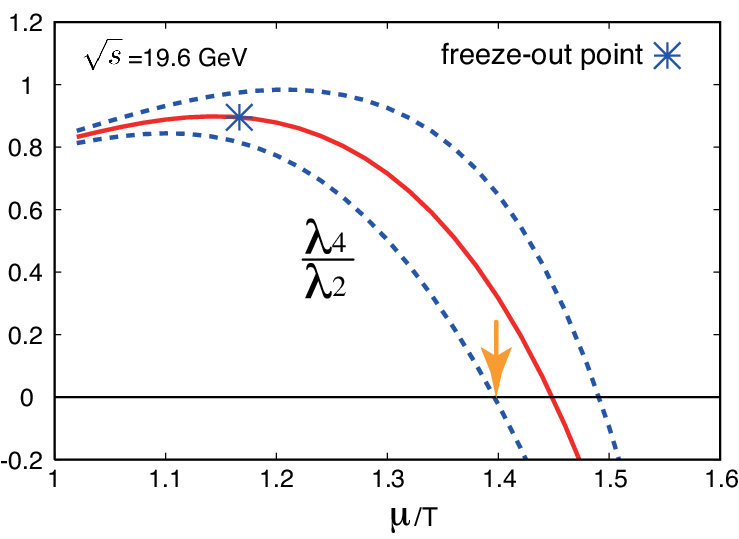}
\caption{
Ratio of the moments $\lambda_4/\lambda_2$
as a function of $\mu/T$ at $\srnn=$19.6 GeV. 
The upper and lower curves are constructed by the same procedure
as used in Fig.\ref{Fig:Sus-27}
}
  \label{Fig:Kurtosis19.6}
\end{figure}

\begin{figure}[h] 
  \centering
  \includegraphics[width=0.8\linewidth]{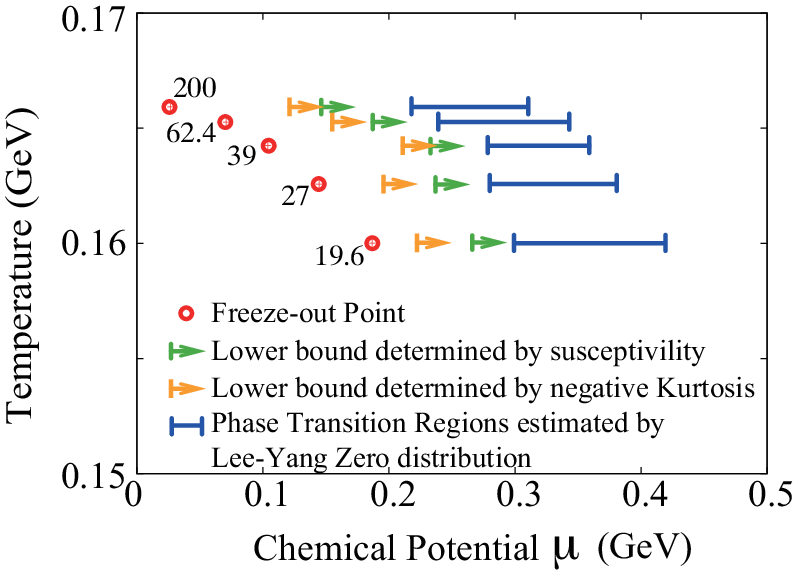}
\caption{
Potential regions of hidden phase transition.
The regions are estimated as the complement of areas where no transition
is evident.
\label{Fig:PhaseBoundary}
}
\end{figure}

Let us suppose that, as $\mu$ increases, 
we encounter a phase transition or a cross-over, and cross it. 
In this case,
we would expect the structure of the moments to be rough in this area. 
At the peak position of the lower curve, indicated by an arrow
in Fig.\ref{Fig:Sus-27}, 
the center line continues to increase. 
We write this value as $\bar{\mu}(T)$, and 
any transition may occur for $\mu \ge \bar{\mu}$.
In other words, this position  $\bar{\mu}$  is a candidate for the lower bound
of the real susceptibility peak.
We then investigate the behavior of  
$\lambda_3/\lambda_2$ and $\lambda_4/\lambda_2$.
Although higher moments have large effects of the finite
$N_{\rm max}$,
they may be a good tool for
detecting the transition region of the QCD phase diagram
\cite{Friman2011}.
The former ratio  does not indicates a significant structure 
for $\mu \le \bar{\mu}$. 
Around the freeze-out points, $\lambda_4/\lambda_2 \sim 1$ 
(Poisson) and it
becomes  negative as $\mu$ increases.
See Fig.\ref{Fig:Kurtosis19.6}.
In Refs.\cite{GavaiGupta2010, Skokov2011,Stephanov2011},
it is argued that
a negative value of $\lambda_4/\lambda_2$ indicates 
that the phase transition has been reached.
We give $\lambda_4/\lambda_2$ at $\sqrt{s_{NN}}=$ 11.5, 19.6,
27, 39, 62.4 and 200 GeV in the appendix.

We take the points where the lower curve of $\lambda_4/\lambda_2$
becomes negative (indicated by an arrow in Fig.\ref{Fig:Kurtosis19.6})
as an another candidate for the lower bound 
$\bar{\mu}(T)$ 
and show them  in Fig.\ref{Fig:PhaseBoundary}.

\section{Lee--Yang zero structure}

Next we extend the fugacity $\xi$ to complex values.
Lee--Yang zeros (LYZs) $\alpha_l$ are zeros 
of the grand partition function $Z(\xi)$
in the complex fugacity plane; that is,
\be
 Z(\xi) = \prod (\xi - \alpha_l).
\label{Eq:LeeYangZero1}
\ee
The distribution of these zeros reflects the phase structure 
of the corresponding statistical system.
Lee and Yang argued that for a finite system,
no zeros appear on the real axis. 
In the thermodynamic limit,
the number of zeros becomes infinite and
the zeros coalesce onto one-dimensional curves \cite{Yang:1952be}.
For the first-order (second-order) phase transition, the coalescing zeros 
cross (pinch) the real positive $\xi$ axis \cite{Stephanov2006},
whereas for the crossover, they do not reach the real axis.

The first pioneering work to calculate the LYZs of the lattice QCD was carried out
by Barbour and Bell \cite{BarbourBell},
who found that the zero distribution is considerably different in the confinement and deconfinement
phases.
In Ref.\cite{FodorKatz2004} the authors calculated the LYZs 
in the complex $\beta=6/g^2$ plane
to distinguish between a crossover and a second-order phase transition 
by investigating the volume dependence.
Ejiri pointed out that this approach is very difficult with finite statistics since
the phase of $\det\Delta(\mu)$ is mixed with the complex $\beta$
\cite{Ejiri2006}.

Although
there have been many phenomenological approaches to extracting 
information on  the QCD phase \cite{Morita2012,Friman2011,Stephanov2011},
no study has yet been carried out  to employ experimental data 
to investigate the QCD phase through the LYZs.

For this purpose
it is important to reliably determine the LYZs;
that is, all zeros must be found without ambiguity, and 
their positions in the complex fugacity plane
must be determined with high accuracy.
We obtain the LYZs as follows.
We first map the problem  to the calculation of the residue of $Z'/Z$: 
\be
{Z'}/{Z} = \sum {1}/({\xi-\alpha_l}).
\label{Eq:dZbyZ}
\ee
The left hand side of Eq.(\ref{Eq:dZbyZ})
is integrated along a contour, and 
the residues inside the contour are summed according to Cauchy's theorem.
Because of  the symmetry $+\mu \leftrightarrow -\mu$, 
if $\alpha_l$ is a solution, 
then $1/\alpha_{l}$ is also a solution. 
Therefore, we only need to search for residues inside
the unit circle.

\begin{figure} 
  \centering
  \includegraphics[width=0.5\linewidth]{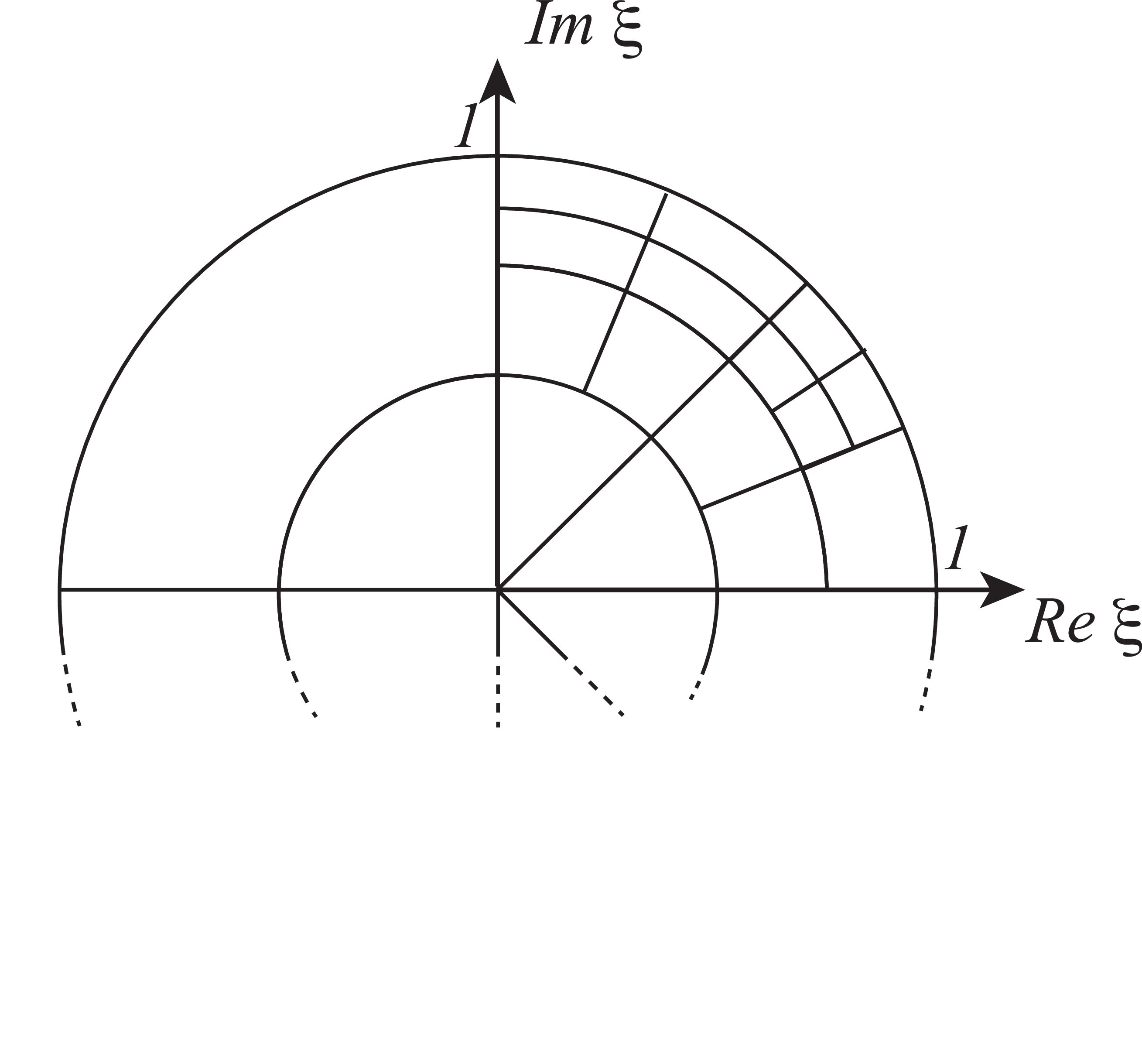}
\caption{
Schematic of the cBK contours in
the divide-and-conquer search for residues.
}
  \label{Fig:CBK}
\end{figure}

Figure \ref{Fig:CBK} shows the
cut Baum-Kuchen (cBK) shape contours used in the study. 
Starting from $0 \le \theta < 2\pi$, and $0 < r \le 1$
in polar coordinates, the region is divided into four pieces,
and the Cauchy integral is evaluated over each section. 
This divide-and-conquer process is iterated as many times
as required (here 20 times). 
When no residue is found in a section, no further divisions
are applied to that section.
At each divide-and-conquer level, we check the conservation of the residue sum.
The technical details of this approach will be presented elsewhere,
including the parallelization.

All calculations were performed using the multiple-precision 
package, FMlib \cite{FMlib}
and the number of significant digits was 50 - 100.
With this algorithm, we can construct LYZ diagrams
from $Z_n$.

\begin{figure} 
  \centering
  \includegraphics[width=0.9\linewidth]{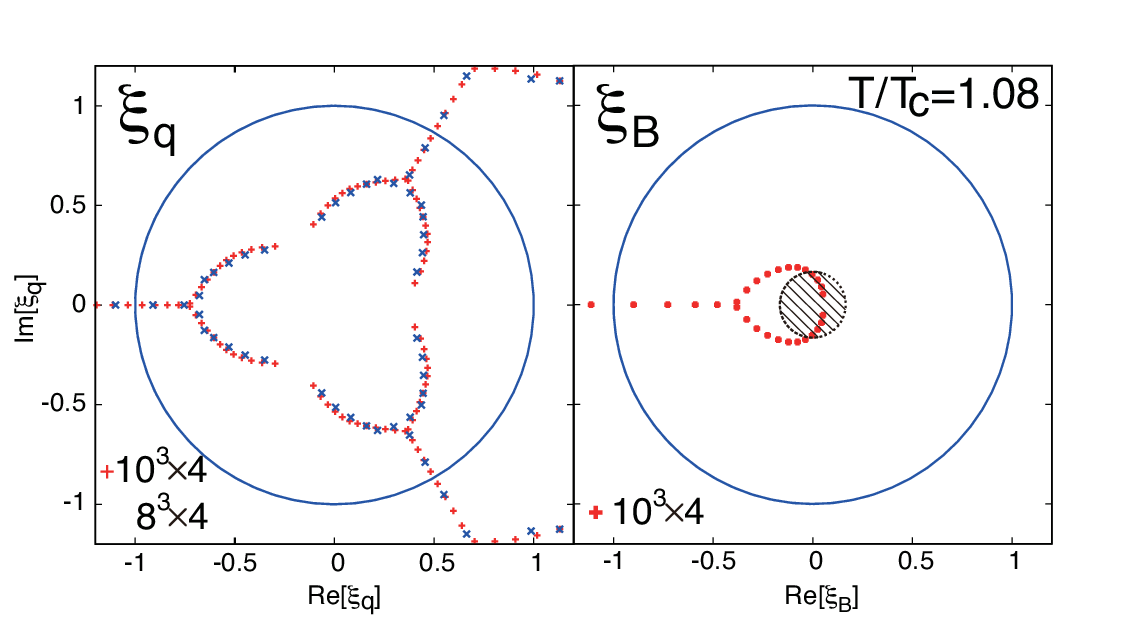}
\caption{
LYZ diagram from the lattice QCD. $T/T_c=1.08$ ($\beta=1.9$).
Left: The lattice size is $10^3\times 4$ and $8^3\times 4$
and $N_{\rm max}=28$ and 14, respectively.
%
%
Right: The same diagram using a lattice size
of $10^3\times 4$, but now it is displaced
in terms of the baryon chemical potential $\xi_B$ instead of the quark
chemical potential $\xi_q$, where $\xi_B =\xi_q^3$.
The shaded circle is an area with high density, i.e.,
$|\xi| \le \exp(-3*0.6)$, where the lattice QCD results have 
limited applicability.
}
  \label{Fig:LYZ-B1950}
\end{figure}

\subsection{Lattice calculations}

First, we study the LYZs obtained by lattice QCD simulation.
Here we do not distinguish between $u$ and $d$ quarks. 
Details of calculating $Z_n$ by lattice QCD simulation
are given in \cite{NN2012},
where the fugacity expansion formula \cite{NN2010} plays an essential role
in obtaining $Z_n$.
%
%
We update 11000 trajectories including 3000 thermalization trajectories.
The measurement is performed every 10 (20) trajectories 
for a $8^3\times4 (10^3\times4)$ lattice size.
A Monte Carlo update is performed with the fermion measure
at $\mu=0$;
thus, we avoid the sign problem due to the complex fermion determinant.
However, an overlap problem may still exist.

\begin{figure} 
  \centering

  \includegraphics[width=0.5\linewidth]{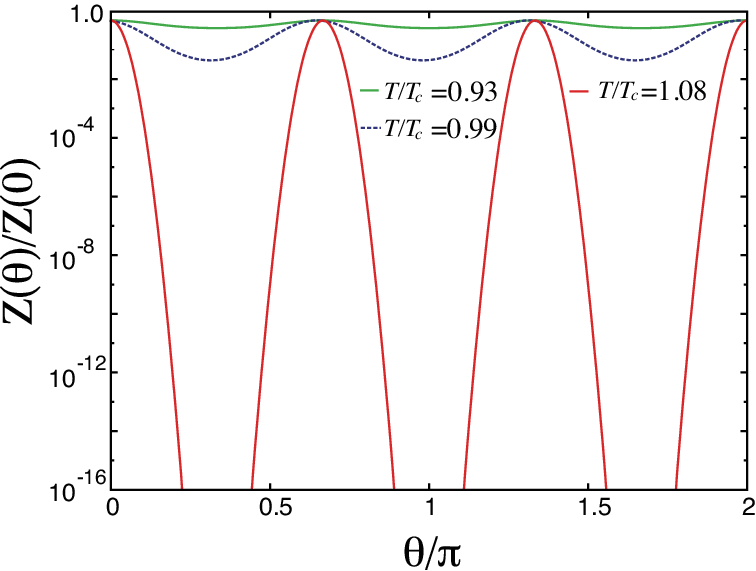}
\caption{
The grand partition function, $Z(\theta)$ as a function of
$\theta\equiv \mbox{Im}(\mu)/T$ for $T/T_c=1.08, 0.99$ and 0.93.
Above $T_c$, $Z(\theta)$ drops rapidly at $\theta=\pi/3, \pi$
and $5\pi/3$, where the free energy 
drops rapidly.
}
  \label{Fig:ZonUnitCircle}
\end{figure}

The LYZ diagram of the lattice QCD  above the phase-transition temperature is
shown in Fig.\ref{Fig:LYZ-B1950}. 
For $Z_n$, we use data 
with  $(\mbox{signal}/\mbox{noise}) \ge 2$.
The condition $Z_n=0$ for $n=1,2\, (\mbox{mod}\, 3)$
is imposed on the lattice data for $Z_n$\cite{Hasenfraz-Toussant,
Kratochvila-Forcrand}
which guarantees $2\pi/3$-translational invariance 
for $\mbox{Im}(\mu)/T$. 
We evaluate the LYZs for $Z(\xi_B)=\sum_{m} Z_{3m} \xi_B^m$ and map the result
onto $\xi=\xi_B^{1/3}$.

In Ref.\cite{NN-JHEP2012} it is shown that
two widely used methods, the multi-parameter reweighting and Taylor expansion,
are consistent for the EoS and
number density up to $\mu_q/T \sim$ 0.8 and for number susceptibility up to  $\mu_q/T \sim$ 0.6.
This implies that the current lattice QCD calculation should not be 
considered reliable in the higher density regions.
In the right panel of Fig.\ref{Fig:LYZ-B1950}, the circle 
corresponding to $|\xi| = \exp(-3\mu_q/T)$ with $\mu_q/T=0.6$ is displayed.

\begin{figure}[ht] 
  \centering
  \includegraphics[width=0.7\linewidth]{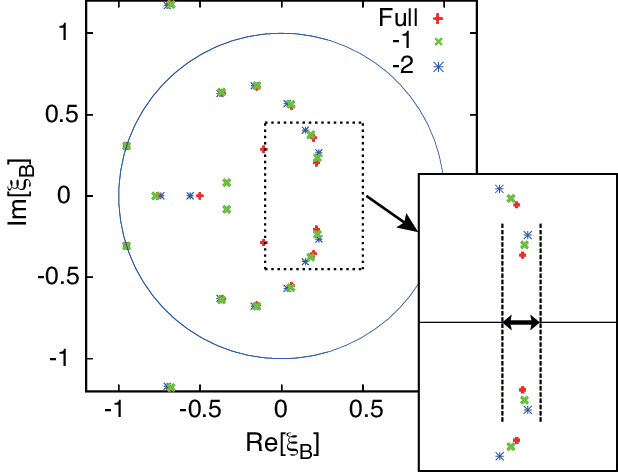}
\caption{
LYZ diagram from RHIC data at  $\srnn=200$ GeV.
``Full'' result is calculated by using all $Z_n$.
``-1'' is calculated by removing a pair with $Z_n$ of $n=\pm N_{\rm max}$.
For ``-2'', we remove further pair of $Z_n$.
The inset is an enlarged view 
near the positive
real axis, which gives the predicted region of the QCD phase transition 
in Fig.\ref{Fig:PhaseBoundary}.
}
  \label{Fig:LYZ-Star200}
\end{figure}

\begin{figure}[ht] 
  \centering
  \includegraphics[width=0.7\linewidth]{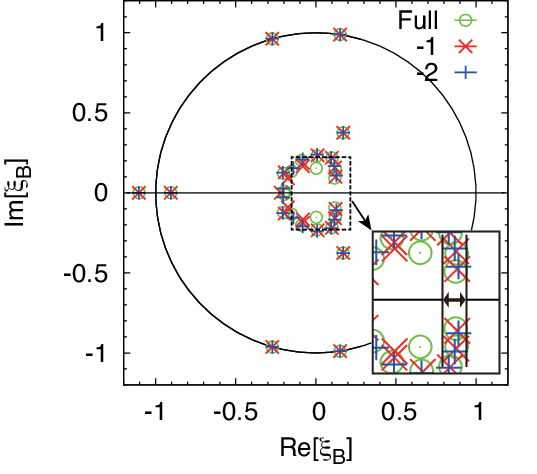}
\caption{
LYZ diagram from RHIC data at $\srnn=19.6$ GeV.
``Full'', 
``-1'' and 
``-2'' denote the same as in Fig.\ref{Fig:LYZ-Star200}
}
  \label{Fig:LYZ-Star19.6}
\end{figure}

Because the $Z_n$ obtained in the confinement
region suffer from significant noise, the LYZ diagram
should be considered qualitative. 
However, despite this,
distinctive differences are observed 
above and below the confinement/deconfinement transition temperature.
At $T \sim 1.2T_c$, 
a line of the zero accumulation appears at $\arg (\xi) = \pm \pi/3$, 
which is consistent with the Roberge-Weiss (RW) phase transition.  
Of course, in order to confirm this is a real RW phase transition, we
must go to large volume (large $N_{max}$) and check that zeros are accumulated.

Roberge and Weiss discussed the regions of pure imaginary chemical potential,
and found that at $\mbox{Im}(\mu)/T=\pi/3 + 2k\pi/3$
($k=0, 1$),
a phase transition occurs for $T\ge T_{\rm RW}$ \cite{RobWei}.
The transition is
the first-order transition above $T_{\rm RW}$, and it is easy to 
detect at experiments.  See  Ref.\cite{DElia-Sanfilippe2009} 
for detailed discussions on the order of the RW phase transition.
See Refs.\cite{Kouno2009,Nagata2015} for more detailed discussion how to occur 
the Roberge-Weisse phase transition.

Using the same lattice setup as that in the present study,
the quantity $T_{\rm RW}$ was estimated to be approximately $1.1 T_c$
\cite{NNImag}. 
This phase transition exerts a clear effect 
on the LYZ diagram 
at $T \sim 1.2T_c$,
while no such effect appears 
at $T \lesssim T_c$.

The fugacity, $\xi$, on the unit circle stands for the pure imaginary chemical potential.
The relation between the pure imaginary, zero and  real chemical potential
are discussed in Refs.\cite{Thies2007, Skokov2011b}.

In the left panel of Fig.\ref{Fig:LYZ-Star200}, we check the volume dependence 
by plotting $8^3\times 4$ and $10^3\times 4$ cases; 
here $N_{max}$ is chosen 
so that $N_{max}/V$ is approximately the same for $V=10^3$ and $8^3$.
In this simulation, $N_{max}/V \sim 1.7\mbox{ fm}^{-3}$, i.e., 
in a cube with one $fm$ on a side, up to  $3\times1.7$ quarks
are included.
See  Ref.\cite{Morita2013}, where  the authors estimate $N_{max}/\sqrt{V}$ 
for obtaining reliable results as a function of the
temperature based on  the quark-meson model. 

Using the relation,
\be
Z = \sum Z_n \xi^n,
\label{Eq:GrandCanonicalSimple}
\ee  
we can see the behavior
of the grand partition function, $Z(\xi=\exp(\mu/T))$, for the 
pure imaginary chemical potential.  We show this behavior in
Fig.\ref{Fig:ZonUnitCircle} for several temperatures.
Lee and Yang pointed out that phase transition regions
correspond to zeros of the grand partition function.
In Fig.\ref{Fig:ZonUnitCircle},
such characteristic behaviors are observed at the Roberge-Weiss
phase transition points in the pure imaginary chemical potential
above $T_c$.
As a consequence, the thermo potential changes rapidly.
Because our $N_{max}$ is 
finite, exact Lee-Yang zeros do not
appear on the pure imaginary chemical potential.
However, rapid decrease of $Z(\theta)$ is seen.
Since 
\be
Z_{n_q}=0 \quad\quad\mbox{for} \quad n_q =1, 2 \quad(\mbox{mod}\quad 3)
\label{Eq:Triality}
\ee
where $n_q$ are the quark number.\footnote{This holds 
both $T>T_c$ and $T<T_c$.}
The symmetry of 
\be
Z(\theta) = Z(\theta+\frac{2\pi}{3})
\ee
leads to the relation (\ref{Eq:Triality}).

Whether zeros appear in Eq.(\ref{Eq:GrandCanonicalSimple})
on the pure imaginary regions, i.e. on the unit circle in
the complex fugacity plane, or not depends on the nature of
$Z_n$.  In other words, whether the Roberge Weise phase transition
occurs or not is the outcome of the dynamics.
In Ref.\cite{Nagata2015}, an explicit example of $Z_n$ that
leads to Roberge Weiss phase transition above $T_c$ is
given and its mechanism is explained.

\subsection{Relativistic Heavy Ion Collisions}

We next consider the LYZ diagrams obtained from RHIC data. 
We construct the grand partition function Eq.(\ref{Eq:GrandCanonical})
for $\srnn=$ 
19.6, 27, 39, 62.4 and 200 GeV
from the data for which $P_n \ge 2$,
and use the cBK method to calculate the LYZ diagrams.
The results for $\srnn=$200 and 19.6 GeV
are shown 
in Fig.\ref{Fig:LYZ-Star200} and \ref{Fig:LYZ-Star19.6}.
To clarify the effect of a finite $N_{\rm max}$, we also calculate the LYZs
while neglecting $Z_n$ for $n=\pm N_{\rm max}$
(``-1'') 
and 
while neglecting $Z_n$ for 
$n=\pm N_{\rm max}$ and $\pm (N_{\rm max}-1)$
(``-2'').

Although some zeros exist on the negative real axis, 
they do not form a line that clearly characterizes
the RW transition, 
which suggests that the data correspond to temperatures
below $T_{\rm RW}$.
Remember that
the net-proton number is not an exactly conserved quantity.
It is therefore very interesting to construct the Lee-Yang zeros and to study their
structure for the conserved charge, such as net charge or strangeness.
 
\begin{figure} 
  \centering
  \includegraphics[width=0.5\linewidth]{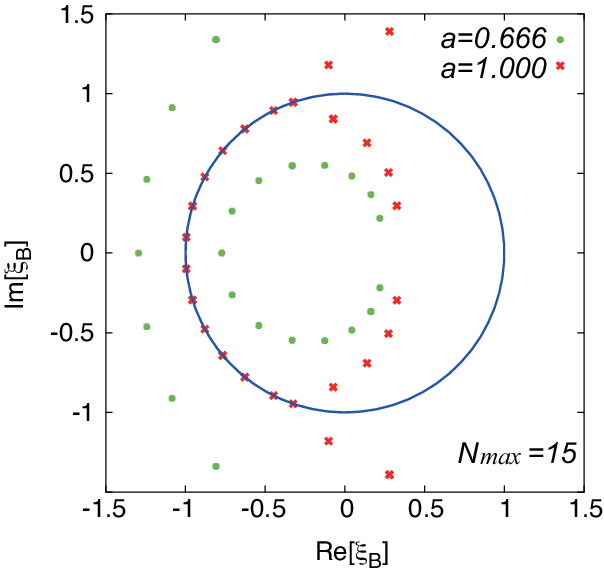}
\caption{
LYZ diagram of the Skellam distribution, in which
  $Z_n$ are given by $I_n(a)$.
  (green) 
$a=6.66$
which corresponds to $\srnn=200$ GeV Star experiment; (red) 
$a=10$.
  In both cases, $N_{\rm max}$ is set to be 15, as in Fig. \ref{Fig:LYZ-Star200} } 
  \label{Fig:LYZ-Skellam}
\end{figure}

In the LYZ diagrams obtained from the RHIC data, no zero appears 
on the positive real-$\xi$ axis.
Possible explanations for this result are that
(i) there is no phase transition, but a crossover occurs at 
these temperatures, 
(ii) the systems are finite, and/or (iii) the $N_{\rm max}$ values 
are too small.
The size of the fireball produced is
comparable to the QCD scale, and thus, explanation (ii) is at least
partially correct.
To further explore the QCD phase transition, 
a larger $N_{\rm max}$ must be attained.

Several LYZ points appear on the unit circle for the all RHIC data.
To understand the meaning of this, we calculate the LYZ diagram for  
the Skellam model.
This is a simple probabilistic model based on
the difference between two Poisson distribution variables $N_1$ and $N_2$, 
and in our case $Z_n=I_{n}(a)$. See Ref.\cite{PBM}.
Here $I_n$ is the modified Bessel function; the parameter $a$ is unique, once the averages of $N_1$ and $N_2$
are given.
In Fig.\ref{Fig:LYZ-Skellam}, the points with $a=0.666$  corresponds to
the multiplicity in the case of $\srnn=200$ GeV.
Two circles appear inside and outside of the unit circle,
and no LYZ point on the unit circle.  
However, if we increase $a$ artificially,
the two circles move closer to the unit circle and several points coalesce on the unit circle.
This indicates that a gross feature of the RHIC multiplicity data 
is configured by the probabilistic origin,
but the LYZ distribution includes additional information on the QCD dynamics.

Finally, we estimate where
the LYZs would intercept or approach 
the positive real axis
as the volume increases, which indicates the QCD phase transition.
These zones are indicated by double -headed arrows 
in the inset of Fig.\ref{Fig:LYZ-Star200}. 
In Fig.\ref{Fig:PhaseBoundary},
the corresponding regions are indicated by 
horizontal lines.
If the multiplicities $P_n$  with larger
$n$ were  to be measured in future experiments, the QCD phase
transition could be pinpointed with more precision. 
Further study of the relation between the baryon and proton number distributions
will improve the analysis here \cite{KitaAsa}.

\section{Concluding Remarks}

A simple but important relation discussed in this paper is
\be
Z(\mu,T) = \sum_{-N_{max}}^{N_{max}} Z_n(T) \xi^n,
\label{Eq:FugacityExpansion}
\ee
e.g., the fugacity expansion of the grand partition function $Z$ with
the coefficients $Z_n$, which are the canonical partition functions.

We have shown how to construct $Z_n$ from experimental net-multiplicities.
In high energy heavy ion central collision, a fireball with $\mu$ and
$T$ is a good picture, at least as a global feature.
Then the selection of an event with a net-multiplicity is a filter
of $Z_n$.

In lattice QCD simulation, we consider the path integral formula
of the grand partition function,
\bea
Z(\mu,T) &=& \int \mathcal{D}U (\det D(\mu))^{N_f} e^{-S_G}
\\  &=& \int \mathcal{D}U \frac{(\det D(\mu))^{N_f}}{(\det D(0))^{N_f}}
(\det D(0))^{N_f} e^{-S_G},
\label{Eq:PathIntegral}
\eea 
where $\det D$, $N_f$ and $S_G$ are fermion determinant, the number of
flavor and gluon action, respectively.
We expand $\det D(\mu)$ in a fugacity series.  Then we get a 
formula (\ref{Eq:FugacityExpansion}).

The canonical partition functions, $Z_n$, depend on only the temperature $T$,
but not on the chemical potential $\mu$.
Therefore once we extract $Z_n$ from $Z(\mu,T)$, we can evaluate $Z$ at 
different values of $\mu'$.

This is very important for exploring the QCD phase boundary:
So far, analyses such as the moments have been done on the
freeze-out points.
But
the freeze-out points realized in the BES experiments 
are in the confinement regions, and not very near to the phase
transition.  
Indeed, Fukushima estimated the baryon density on the freeze-out
line using the hadron resonance gas model, and found that the maximum
of the baryon density is realized at $\sqrt{s_{NN}}=8$ GeV, but
its value is only slightly larger than the normal nuclear density 
$\rho_0$\cite{FukushimaHRG}.
Using our formula, (\ref{Eq:FugacityExpansion}), we can probe higher
chemical potential regions from $Z_n$ constructed at the chemical
freeze-out point.

We calculated the moments (\ref{Eq:MomentsDef}), and 
saw their behavior when increasing $\mu/T$.
We checked the effects of the finite $N_{max}$.

The formula (\ref{Eq:FugacityExpansion}) makes it possible to calculate
$Z(\mu,T)$ at complex values of $\mu$.
We studied the Lee-Yang zero distribution data of both the experiments
and the lattice simulations.
To our knowledge, this is the first Lee-Yang zero analysis of the RHIC data.
Lattice results told us that the zeros corresponding to 
Roberge-Weiss transitions appear above $T_c$.  We did not see
such structure in the LYZ of RHIC data.  
This suggests that these experimental
data are produced below $T_{\rm RW}$, where $T_{\rm RW}$ is Roberge-Weiss 
transition temperature that is around $1.1 T_c$.

There were claims that the RW transition formulated in the Euclidean space
can not be interpreted as a physics object in the Minkowski space
by considering its cosmological consequences
\cite{Smilga1994,Belyaev1992}.
In this paper, we discussed the relation between the RW phase transition 
and the high energy heavy ion experiment, which hopefully 
advance understanding of the QCD phase diagram
\footnote{
The authors thank the referee to point out the argument.}
.

The approach investigated here is based on the simple statistical mechanics,
and is easy to use for extracting information from experimental data.
It
works equally in analyzing experimental data and lattice QCD simulations.
In the latter case, we can study the real chemical potential regions without
Taylor expansions.

Several problems that should be clarified in future are
\begin{itemize}
\item
We can study finite real chemical potential regions 
in lattice QCD simulation and
the sign problem does not appear here.
A possible obstacle is the overlap problem.  
In our lattice Monte Carlo simulations, the gauge configurations are
produced at $\mu=0$ as in Eq.(\ref{Eq:PathIntegral}).
Such configurations may not overlap enough with states of large $n$.
The canonical partition functions $Z_n$ are given in
\be
Z_n = \int_{-\pi}^{+\pi} \frac{d\theta}{2\pi}Z(\theta\equiv \mu_I/T) e^{i\theta n},
\ee
where $\mu_I$ is pure imaginary chemical potential. 
In other words, $Z(\xi)$ on the unit circle has whole information
\footnote{
This is of course under the condition that the number operator $\hat{N}$
is a good quantum number.  For regions of the color-flavor-locked phase,
for example, the number operator is anymore a good quantum number,
and the canonical formulation does not work properly.
}
.
On this circle, $\det D$ is real, and the Monte Carlo simulation is possible.
Therefore, in addition to $\mu=0$, more points on the unit circle
may improve the overlap problem.

\item
Since $(\det D(\mu))^2$ is real positive on the unit circle in $\xi$ plane,
$Z(\xi)$ cannot be zero.  Therefore Lee-Yang zeros on the unit circle
are artifact of $N_{max}$\footnote{We thank K. Splittorf for bringing out
our attention to this point.}, or the $u-$ and $d-$ quark contribution,
$\det D(\mu_u)\det D(\mu_d)$ gives such a effect. 
Therefore we need larger $N_{max}$ both in experiments and lattice QCD calculations.

\item
It is difficult to measure experimentally the net baryon multiplicity.
One possible approach is to study the difference between net-proton multiplicity
and net-baryon multiplicity.

\item
Net strangeness and net charge multiplicity are analyzed in the same way.
\item
The relation between the order of the RW transition and the Lee-Yang zeros
should be studied more quantitatively near the end-point.
The volume dependence of Lee-Yang zeros in the vicinity of a transition point depends on the order of the transition.
If the transition is the second order, Lee-Yang zeros approach 
the RW phase points
according to the volume dependence with a critical exponent.

\item
The Lee-Yang zeros calculated in the lattice QCD simulation at high temperature
suggest the RW transition.  But there might be danger to misidentify 
the thermal singularity with the Lee-Yang zeros.  The former
appears as a branch point at $\xi= e^{-m_N/T}$ inside the unit circle 
and a cut to $-\infty$ on real negative $\xi$ axis
\cite{Thies2007,Skokov2011}.
In order to avoid such misidentification, it is important to check the
volume dependence.
\item
The Lee-Yang zeros are closing to the real positive axis if the system
indicates the phase transition.  Therefore it is interesting the volume
dependence of the Lee-Yang zeros.  Such information may be obtained
by varying the atomic number of beam/target.
\item
Recently, based on the canonical partition functions, detailed analyses 
of effects of $N_{max}$ on the Lee-Yang zeros and the RW transition
were reported,
using the random matrix model and the  saddle point approximation
\cite{Morita-PRD92-114507,Kashiwa-PRD91-094507}.
It is valuable to 
repeat calculations presented in this paper with referring these 
analyses.
%
\item
Both experimental and lattice QCD data include errors.  It is important to check
whether obtained Lee-Yang zeros are stable or not against these errors,
since it requires extreme care to calculate zeros of high order polynomials.
In lattice QCD, we have investigated statistical errors of Lee-Yang zeros caused by 
statistical error of $Z_n$ by using a bootstrap analysis. 
We found 
that the Lee-Yang zeros of high temperature QCD are stable 
near the unit circle on the fugacity plane \cite{Nagata2015}.
If errors of $P_n$ are available, we can investigate the error of Lee-Yang zeros 
for experimental data in a similar manner to lattice QCD. 
In Ref. \cite{Halasz2000}, the authors addressed the numerical error of Lee-Yang zeros using a random matrix model
with finite chemical potential.  
They calculated zeros of the grand partition function in the complex $\mu$ and observed
their behavior when they change the exact polynomial coefficients, $c_k$ to $\tilde{c}_k$
as $\tilde{c}_k = c_k (1+R_k \epsilon)$ .
Here $R_k$ are random numbers between $[-1,1]$.
They found that the deviation of the zeros on the real $\mu$ axis from the exact values
is approximately proportional to $\log\epsilon$, and that the structure os Lee-Yang zeros
is not spoiled.
%
\end{itemize}
 
\section*{Acknowledgment \\}

This work grew out of a stimulus provided to one of the authors (A.N.)
by L. McLerran and N. Xu
at the `QCD Structure' workshop in Wuhan. 
We thank 
N. Xu,  X. Luo,  C. Sasaki, K. Shigaki, M. Kitazawa and  V. Skokov 
for valuable discussions. 
We are indebted to Ph. de Forcrand and K. Morita
for critical reading of the manuscript and valuable comments. 
We wish to thank S. Aoki, T. Hatsuda, K. Redlich and M. Yahiro 
for their continuous interest and encouragement. 
The work was completed after very stimulating
discussion with B. Friman, J. Knoll, V. Koch, K. Morita
and J. Wambach at GSI.
The calculations were performed on  SX-9, SX-ACE and SAHO, at RCNP Osaka, 
RICC at Riken and SR16000  at KEK.
This work was supported by
Grants-in-Aid for Scientific Research
20105003-A02-0001, 23654092 and 24340054.


\appendix

\section{Canonical partition functions, $Z_n$ for RHIC data}

In this appendix, we give $Z_n$ obtained in Sec.\ref{Sec:Zn}.
Since $Z_{-n}=Z_n$, we show only $n \ge 0$ values.
The data are normalized as $Z_0=1$ at each energy.
$\Delta Z_n$ are estimated errors.

\begin{table}[ht]
\begin{minipage}{0.45\hsize}
\begin{tabular}{rcc}
\hline 
$n$ &  $Z_n$  & $\Delta Z_n$  \\
\hline 
  0  &  0.10000E+01  &  0.62112E-07  \\
  1  &  0.62361E+00  &  0.69741E-07  \\
  2  &  0.33222E+00  &  0.66895E-07  \\
  3  &  0.12597E+00  &  0.45672E-07  \\
  4  &  0.40292E-01  &  0.26301E-07  \\
  5  &  0.10419E-01  &  0.12245E-07  \\
  6  &  0.23949E-02  &  0.50681E-08  \\
  7  &  0.46882E-03  &  0.17863E-08  \\
  8  &  0.83546E-04  &  0.57315E-09  \\
  9  &  0.13110E-04  &  0.16194E-09  \\
 10  &  0.18749E-05  &  0.41698E-10  \\
 11  &  0.24529E-06  &  0.98222E-11  \\
 12  &  0.29522E-07  &  0.21285E-11  \\
 13  &  0.32786E-08  &  0.42561E-12  \\
 14  &  0.33756E-09  &  0.78901E-13  \\
 15  &  0.33251E-10  &  0.13994E-13  \\
 16  &  0.30000E-11  &  0.22732E-14  \\
 17  &  0.26163E-12  &  0.35694E-15  \\
 18  &  0.21094E-13  &  0.51818E-16  \\
 19  &  0.15816E-14  &  0.69954E-17  \\
 20  &  0.11258E-15  &  0.89653E-18  \\
 21  &  0.83712E-17  &  0.12003E-18  \\
 22  &  0.56668E-18  &  0.14630E-19  \\
 23  &  0.38519E-19  &  0.17905E-20  \\
 24  &  0.18180E-20  &  0.15215E-21  \\
 25  &  0.70247E-22  &  0.10585E-22  \\
 26  &  0.82136E-23  &  0.22285E-23  \\
 27  &  0.31360E-24  &  0.15320E-24  \\
 28  &  0.34922E-25  &  0.30717E-25  \\
\hline
\hline
\end{tabular}
\caption{$\sqrt{s_{NN}}$=11.5 GeV}
\end{minipage}
\begin{minipage}{0.45\hsize}
\begin{tabular}{rcc}
\hline 
$n$ &  $Z_n$  & $\Delta Z_n$  \\
\hline 
0 & 0.10000E+01 & 0.11360E-01 \\
1 & 0.88457E+00 & 0.16150E-01 \\
2 & 0.65521E+00 & 0.19945E-01 \\
3 & 0.40203E+00 & 0.19813E-01 \\
4 & 0.20550E+00 & 0.14453E-01 \\
5 & 0.90348E-01 & 0.11442E-01 \\
6 & 0.34569E-01 & 0.70907E-02 \\
7 & 0.11665E-01 & 0.38755E-02 \\
8 & 0.35214E-02 & 0.18949E-02 \\
9 & 0.96801E-03 & 0.84371E-03 \\
10 & 0.23849E-03 & 0.33668E-03 \\
11 & 0.54473E-04 & 0.12456E-03 \\
12 & 0.11384E-04 & 0.42160E-04 \\
13 & 0.22056E-05 & 0.13231E-04 \\
14 & 0.40106E-06 & 0.38968E-05 \\
15 & 0.67442E-07 & 0.10614E-05 \\
16 & 0.10994E-07 & 0.28023E-06 \\
17 & 0.16066E-08 & 0.66333E-07 \\
18 & 0.21451E-09 & 0.14345E-07 \\
19 & 0.27632E-10 & 0.29929E-08 \\
20 & 0.39570E-11 & 0.69420E-09 \\
21 & 0.46594E-12 & 0.13240E-09 \\
22 & 0.46170E-13 & 0.21250E-10 \\
23 & 0.57645E-14 & 0.42973E-11 \\
\hline
\end{tabular}
\caption{$\sqrt{s_{NN}}$=19.6 GeV}
\end{minipage}
\end{table}
\begin{table}[ht]
\begin{minipage}{0.45\hsize}
\begin{tabular}{rcc}
\hline 
$n$ &  $Z_n$  & $\Delta Z_n$  \\
\hline 
0 & 0.10000E+01 & 0.55468E-02 \\
1 & 0.90362E+00 & 0.83594E-02 \\
2 & 0.68168E+00 & 0.10515E-01 \\
3 & 0.43157E+00 & 0.10829E-01 \\
4 & 0.23245E+00 & 0.95400E-02 \\
5 & 0.10813E+00 & 0.73858E-02 \\
6 & 0.44216E-01 & 0.58895E-02 \\
7 & 0.15962E-01 & 0.30883E-02 \\
8 & 0.52022E-02 & 0.16718E-02 \\
9 & 0.15312E-02 & 0.81733E-03 \\
10 & 0.41235E-03 & 0.36560E-03 \\
11 & 0.10014E-03 & 0.14748E-03 \\
12 & 0.22725E-04 & 0.55589E-04 \\
13 & 0.47693E-05 & 0.19378E-04 \\
14 & 0.93667E-06 & 0.63214E-05 \\
15 & 0.16794E-06 & 0.18826E-05 \\
16 & 0.30799E-07 & 0.57346E-06 \\
17 & 0.47106E-08 & 0.14569E-06 \\
18 & 0.69913E-09 & 0.35914E-07 \\
19 & 0.11524E-09 & 0.98328E-08 \\
20 & 0.93232E-11 & 0.13213E-08 \\
21 & 0.17834E-11 & 0.41984E-09 \\
\hline
\end{tabular}
\caption{$\sqrt{s_{NN}}$=27 GeV}
\end{minipage}
\begin{minipage}{0.45\hsize}
\begin{tabular}{rcc}
\hline 
$n$ &  $Z_n$  & $\Delta Z_n$  \\
\hline 
0 & 0.10000E+01 & 0.20603E-02 \\
1 & 0.91388E+00 & 0.29289E-02 \\
2 & 0.69779E+00 & 0.34996E-02 \\
3 & 0.45149E+00 & 0.35006E-02 \\
4 & 0.25020E+00 & 0.30516E-02 \\
5 & 0.12024E+00 & 0.22448E-02 \\
6 & 0.50805E-01 & 0.14914E-02 \\
7 & 0.19122E-01 & 0.93227E-03 \\
8 & 0.64636E-02 & 0.51254E-03 \\
9 & 0.19762E-02 & 0.22024E-03 \\
10 & 0.54891E-03 & 0.95311E-04 \\
11 & 0.14064E-03 & 0.38048E-04 \\
12 & 0.32779E-04 & 0.13816E-04 \\
13 & 0.73396E-05 & 0.48197E-05 \\
14 & 0.15625E-05 & 0.15986E-05 \\
15 & 0.28234E-06 & 0.45005E-06 \\
16 & 0.50870E-07 & 0.12634E-06 \\
17 & 0.10789E-07 & 0.41746E-07 \\
18 & 0.18428E-08 & 0.11109E-07 \\
19 & 0.12762E-09 & 0.11987E-08 \\
20 & 0.33882E-10 & 0.49581E-09 \\
21 & 0.11993E-10 & 0.27344E-09 \\
\hline
\end{tabular}
\caption{$\sqrt{s_{NN}}$=39 GeV}
\end{minipage}
\end{table}

\begin{table}[ht]
\begin{minipage}{0.45\hsize}
\begin{tabular}{rcc}
\hline 
$n$ &  $Z_n$  & $\Delta Z_n$  \\
\hline 
0 & 0.10000E+01 & 0.47305E-03 \\
1 & 0.92168E+00 & 0.79653E-03 \\
2 & 0.72216E+00 & 0.11385E-02 \\
3 & 0.48191E+00 & 0.13929E-02 \\
4 & 0.27792E+00 & 0.14614E-02 \\
5 & 0.14035E+00 & 0.13333E-02 \\
6 & 0.62429E-01 & 0.11101E-02 \\
7 & 0.24831E-01 & 0.78510E-03 \\
8 & 0.89221E-02 & 0.51745E-03 \\
9 & 0.28887E-02 & 0.33828E-03 \\
10 & 0.86915E-03 & 0.15845E-03 \\
11 & 0.23370E-03 & 0.83409E-04 \\
12 & 0.60694E-04 & 0.39564E-04 \\
13 & 0.14052E-04 & 0.16730E-04 \\
14 & 0.30132E-05 & 0.65523E-05 \\
15 & 0.56945E-06 & 0.22616E-05 \\
16 & 0.10396E-06 & 0.75408E-06 \\
17 & 0.26627E-07 & 0.35277E-06 \\
18 & 0.47347E-08 & 0.11457E-06 \\
\hline
\end{tabular}
\caption{$\sqrt{s_{NN}}$=62.4 GeV}
\end{minipage}
\begin{minipage}{0.45\hsize}
\begin{tabular}{rcc}
\hline 
$n$ &  $Z_n$  & $\Delta Z_n$  \\
\hline 
 0 & 0.10000E+01 & 0.16546E-03 \\
 1 & 0.92030E+00 & 0.26094E-03 \\
 2 & 0.72070E+00 & 0.35016E-03 \\
 3 & 0.48388E+00 & 0.40286E-03 \\
 4 & 0.28066E+00 & 0.40041E-03 \\
 5 & 0.14265E+00 & 0.34875E-03 \\
 6 & 0.63967E-01 & 0.26797E-03 \\
 7 & 0.25682E-01 & 0.18436E-03 \\
 8 & 0.93739E-02 & 0.11531E-03 \\
 9 & 0.30591E-02 & 0.64484E-04 \\
10 & 0.95799E-03 & 0.34604E-04 \\
11 & 0.24394E-03 & 0.15099E-04 \\
12 & 0.57285E-04 & 0.60760E-05 \\
13 & 0.18142E-04 & 0.32974E-05 \\
14 & 0.31923E-05 & 0.99425E-06 \\
15 & 0.92947E-06 & 0.49606E-06 \\
\hline
\end{tabular}
\caption{$\sqrt{s_{NN}}$=200 GeV}
\end{minipage}
\end{table}

\section{Moments $\lambda_4/\lambda_2$ for RHIC data}

\begin{figure}
  \centering
  \includegraphics[width=0.9\linewidth]{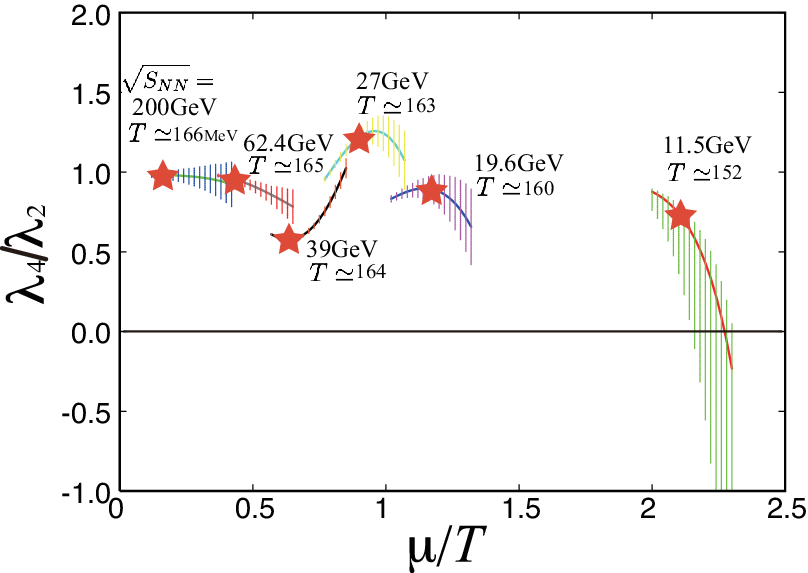}
  \caption{
The ratio of the moments $\lambda_4$ and $\lambda_2$, which
correspond $\kappa \sigma^2/T^2$ as a function of $\mu/T$
for RHIC beam energy scan data at
$\sqrt{s_{NN}}=$11.5, 19.6, 27, 39, 62.4 and 200 GeV.  
Here $\kappa$ and $\sigma^2$ are 
the kurtosis and the variance, respectively.
The red starts indicate the freeze-out points.
}
\end{figure}

\end{document}